\documentclass[fleqn,twoside]{article}
\usepackage{espcrc2}

\newcommand{\sla}[1]{{\ooalign{\hfil/\hfil\crcr$#1$}}}
\newcommand{\f}[2]{\frac{#1}{#2}}

\newcommand{\lb}[0]{\l[}  % abbrev.of left bracket
\newcommand{\rb}[0]{\r]}
\newcommand{\lc}[0]{\l\{} % abbrev.of left curly bracket
\newcommand{\rc}[0]{\r\}}

\newcommand{\be}{\begin{eqnarray}}
\newcommand{\ee}{\end{eqnarray}}
\newcommand{\bc}{\begin{center}}
\newcommand{\ec}{\end{center}}

\def\l{\left}
\def\r{\right}

\def\ga{\gamma}
\def\g5{\gamma_5}

\title{Properties of a New Class of Lattice Dirac
Operators}

\author{Kazuo Fujikawa\address[Tokyo]{Department
of Physics, University of Tokyo, Bunkyo-ku, 
Tokyo 113, Japan} 
and 
Masato Ishibashi\addressmark[Tokyo]\thanks{
Talk presented by M. Ishibashi}}

\begin{document}
\begin{abstract}
A new class of lattice Dirac operators $D$ have been recently
proposed on the basis of the generalized Ginsparg-Wilson 
relation, $\gamma_5(\gamma_5 D) + (\gamma_5 D)\gamma_5
=2a^{2k+1}(\gamma_5 D)^{2k+2}$, where $k$ is a non-negative
integer. We discuss the index theorem and locality properties
for this general class of lattice Dirac operators.
\vspace{1pc}
\end{abstract}

\maketitle

\section{Introduction}
We have recently witnessed a remarkable progress 
in the treatment of lattice fermions. A basic algebraic
relation which lattice Dirac operators should satisfy 
was clearly stated by Ginsparg and Wilson\cite{GW},
and an explicit solution to the algebra was 
given\cite{N1}.
This solution exhibits quite interesting chiral 
properties\cite{H}\cite{HLN}\cite{L1}, including 
locality 
properties\cite{HJL}\cite{N2}. In the mean 
time, a new class
of lattice Dirac operators $D$ have been proposed on the 
basis of the algebraic relation\cite{F1}
\begin{eqnarray}
\label{ggw}
\g5 D + D\g5 =2aD(a\g5 D)^{2k}\g5 D
\end{eqnarray}
where $k$ stands for a non-negative integer, and $k=0$ corresponds
to the ordinary Ginsparg-Wilson relation\cite{GW}. 
In the following sections we discuss 
the properties of the operators satisfying this 
general Ginsparg-Wilson relation. 
 
\section{A new class of lattice Dirac operators}
We first define
\begin{eqnarray}
\label{gwo1}
&&H_{(2k+1)}\nonumber\\
&&\equiv\f{1}{2}\g5\left(1+ D_W^{(2k+1)} 
\f{1}{\sqrt{D^{(2k+1)\dagger}_W D_W^{(2k+1)}}}
\right)\nonumber\\
\end{eqnarray}
where
\begin{eqnarray}
D_W^{(2k+1)}&=&i(\sla{C})^{2k+1} + (B)^{2k+1} - 
\left(\f{m_0}{a}\right)^{2k+1}\nonumber\\
\end{eqnarray}
where $C_\mu$ is the covariant difference operator
and $B$ is the covariant Wilson term.
%C_{\mu}(x,y)&=&\frac{1}{2a}[\delta_{x+\hat{\mu} a,y}
%U_{\mu}
%(y)-\delta_{x,y+\hat{\mu} a}U^{\dagger}_{\mu}(x)],
%\nonumber\\
%B(x,y)&=&\frac{r}{2a}\sum_{\mu}[2\delta_{x,y}-
%\delta_{y+\hat{\mu} a,x}U_{\mu}^{\dagger}(x)
%\nonumber\\
%&&-\delta_{y,x+\hat{\mu} a}U_{\mu}(y)],
%\nonumber\\
%U_{\mu}(y)&=& \exp [iagA_{\mu}(y)],
%\end{eqnarray}
The solution $D$ of the general Ginsparg-Wilson 
relation(\ref{ggw}) 
is then given by
\begin{eqnarray}
\label{gwo2}
D&=&\frac{1}{a}\g5\left(H_{(2k+1)}\right)^{1/(2k+1)}
\end{eqnarray}
where $k=0$ corresponds to the overlap Dirac 
operator\cite{N1}.
When the parameter $m_0$ is chosen as $0<m_0<2r$ in these
operators,
the free propagators have a single massless pole and are free
from species doublers. We next see that the case 
of $k\ge 1$ is
better than $k=0$ on the scaling property for $a\to 0$.
For the operators(\ref{gwo2}), in the near continuum configurations
we see the behaviours as follows,
\begin{eqnarray}
D&\simeq& i\sla{D}+ a^{2k+1}(\g5 i\sla{D})^{2k+2}
\end{eqnarray}
where $\sla{D}= \ga_\mu(\partial_\mu+igA_\mu)$.
The first terms in these expressions stand for the leading terms
in chiral symmetric terms, and the second terms stand for the
leading terms in chiral symmetry breaking terms. This shows
that one can improve the chiral symmetry for larger $k$. 
As another manifestation of this property, the spectrum 
of 
the operators with $k\ge 1$ is closer to that of 
the continuum operator
in the sense that the small eigenvalues of $D$ accumulate along
the imaginary axis (which is a 
result of taking a $2k+1$-th root)\cite{C}, compared 
to the overlap Dirac operator for which the eigenvalues of
$D$ draw a perfect circle in the complex eigenvalue plane.  

\section{Index theorem}
We first examine the lattice chiral symmetry\cite{L1}. 
The fermion action 
which has the generalized operators(\ref{gwo2}) 
is invariant
under the global transformation:
\begin{eqnarray}
\psi&\rightarrow& \psi^\prime = \psi + i\epsilon\Gamma_5\psi,
\nonumber\\
\bar{\psi}&\rightarrow& \bar{\psi}^\prime =\bar{\psi} + 
i\epsilon\bar{\psi}
\g5 \Gamma_5\g5,
\end{eqnarray} 
where $\Gamma_5 = \g5-(a\g5 D)^{2k+1}$.
But the fermion measure produces the Jacobian factor:
\begin{eqnarray}
{\cal D}\bar{\psi}^\prime{\cal D}\psi^\prime&=& 
J{\cal D}\bar{\psi}{\cal D}\psi,\nonumber\\
J&=&\> exp\left(-2i\epsilon Tr 
\Gamma_5\right).
\end{eqnarray}   
In this expression $Tr \Gamma_5$ is regarded as 
the chiral anomaly on the lattice.
We next analyze the index relation for the generalized Dirac
operators. 
We consider the eigenvalues $\lambda_n$ and the eigenmodes
$\phi_n$ of the hermitian operator $\g5 D$.
The zero modes are defined by
\begin{eqnarray}
(\g5 D) \phi_0 = 0.
\end{eqnarray}
Using $\Gamma_5(\g5 D) + (\g5 D)\Gamma_5 =0$ from the 
general algebra, we obtain
\begin{eqnarray}
(\g5 D) (\g5\phi_0) = 0.
\end{eqnarray} 
Therefore we can assign the chiralities for zero modes:
\begin{eqnarray}
\g5 \phi_0^{(\pm)}=\pm \phi_0^{(\pm)}
\end{eqnarray}
The other modes ($\lambda_n \neq 0$) are defined by
\begin{eqnarray}
(\g5 D)\phi_n&=& \lambda_n\phi_n,\nonumber\\  
(\g5 D)\Gamma_5\phi_n&=& -\lambda_n\Gamma_5\phi_n,  
\end{eqnarray}
And then we obtain $\left(\phi_n,\Gamma_5\phi_n\right) =0$
 for $\lambda_n\neq 0$.
 From the above analysis 
$Tr \Gamma_5$ is written as follows ($n=0$ stands for 
$\lambda_n=0$),
\begin{eqnarray}
Tr \Gamma_5 &\equiv& \sum_{n}
\left(\phi_n,\Gamma_5
\phi_n\right)\nonumber\\
&=&\sum_{n=0}\left(\phi_n,\Gamma_5\phi_n\right)+
\sum_{n\neq 0}\left(\phi_n,\Gamma_5\phi_n\right)
\nonumber\\
&=&\sum_{n=0}\left(\phi_n,\Gamma_5\phi_n\right)\nonumber\\
&=&\sum_{n=0}\left(\phi_n,(\g5 - (\g5 aD)^{2k+1})
\phi_n\right)\nonumber\\
&=&\sum_{n=0}\left(\phi_n,\g5\phi_n\right)\nonumber\\
&=&n_+ - n_- =\, index\nonumber
\end{eqnarray}
This is the index theorem on the lattice. 
Next we evaluate
$Tr \Gamma_5$ for $a\to 0$\cite{FI1}.
The local version of the trace is
\begin{eqnarray}
&&tr\Gamma_{5}(x)=tr[\gamma_{5}-(\gamma_{5}aD)^{2k+1}]
\nonumber\\
&&=-tr\frac{1}{2}\gamma_{5}[D_{W}^{(2k+1)}\frac{1}
{\sqrt{(D_{W}^{(2k+1)})^{\dagger}D_{W}^{(2k+1)}}}]\nonumber
\end{eqnarray}
In this expression $D_{W}^{(2k+1)}$ includes the lattice 
spacing $a$. Then we expand the above expression in $a$ 
and take $a\to 0$. After the straightforward 
calculations, we obtain
\begin{eqnarray}
tr \Gamma_5(x)= I_{2k+1}(r,m_0)
g^2\> tr\epsilon^{\mu\nu\rho\sigma}
F_{\mu\nu}F_{\rho\sigma}.\nonumber
\end{eqnarray}
We can show that $I_{2k+1}(r,m_0)$ is independent of
$k,r$ and $m_0$ and that $I_{2k+1}(r,m_0)=1/32\pi^2$.
Therefore we recover the index theorem in the continuum 
theory:
\begin{eqnarray}
n_+ - n_- = \int d^4 x \f{g^2}{32\pi^2} 
tr\epsilon^{\mu\nu\rho\sigma}
F_{\mu\nu}F_{\rho\sigma}.\nonumber
\end{eqnarray}  
in the naive continuum limit for any 
operator in (\ref{gwo2}).
\section{Locality property}
In this section we discuss the locality property
\cite{FI2}.
We first see the locality of the free operator. 
The generalized Dirac operators in the free 
fermion case are
written as
\begin{eqnarray}
\label{fh}
&&H(p)=\g5\left(\f{1}{2}\right)^{\f{k+1}{2k+1}}
\left(\f{1}{\sqrt{F_k}}\right)^{\f{k+1}{2k+1}}\nonumber\\
&&\times \lc \left(\sqrt{F_k}+\tilde{M}_k
\right)^{\f{k+1}{2k+1}}-\left(\sqrt{F_k}-\tilde{M}_k
\right)^{\f{k}{2k+1}}\sla{s}\rc,\nonumber\\
\end{eqnarray}
where
\begin{eqnarray}
&&\tilde{M}_k = \lb \sum_{\mu}(1-c_\mu)\rb^{2k+1}
-1\nonumber\\
&&F_k=(s^2)^{2k+1}+\tilde{M}_k^2\nonumber\\
&&c_\mu=\cos ap_\mu,\> \sla{s}=\sum_{\mu}\ga_\mu\sin 
ap_\mu,\nonumber\\
&&s^2=\sum_{\mu}(\sin ap_\mu)^2.\nonumber
\end{eqnarray}
Here $H=a\g5 D$ and we set $r=1$ and $m_0=1$.
Now we want to know if $H$ is exponentially local.
For this purpose we note the next statement:
If $H(p)$ is differentiable for infinite times with respect
to $p$, $H(x,y)$ is exponentially local\cite{K}.
Considering the one-dimensional case for simplify, this
proof goes as follows,
\begin{eqnarray}
\f{\partial}{\partial p^l}\,H(p)&=&\f{\partial}{\partial p^l}
\int dx\,e^{ipx}H(x)\nonumber\\
&=&\int dx(ix)^l H(x)e^{ipx}<\infty\nonumber
\end{eqnarray}
This last inequality for any $l$ implies
\begin{eqnarray}
\| H(x) \| < C\,e^{-\theta x},\qquad \theta >0.\nonumber 
\end{eqnarray}
Now all we need to do is to know if $H(p)$ is 
infinite times differentiable. From the expression
of $H(p)$(\ref{fh}), the infinite times
differentiablity means that the following terms, 
\begin{eqnarray}
&&\left(\f{1}{\sqrt{F_k}}
\right)^{\f{(k+1)}{(2k+1)}}, 
\left(\sqrt{F_k}+\tilde{M}_k
\right)^{\f{(k+1)}{(2k+1)}},\nonumber\\
&&\left(\sqrt{F_k}-\tilde{M}_k
\right)^{\f{k}{(2k+1)}},
\end{eqnarray}
are infinite times differentiable.
Noting that there is a mass gap in $F_k$, 
we can prove this\cite{FI2}. Therefore 
the free operator $H$ is local. We also
presented a crude estimate of the localization length, examining 
the singularity of 
the Dirac operators. This suggests
that the operators with gauge fields are local
for sufficiently weak background gauge fields.
On the other hand,
it is difficult to see the locality of
the interacting operator $H$ directly. But it
may be true that if $H_{(2k+1)}=H^{(2k+1)}$ is 
local, $H$ is also local. In
the one-dimensional integral, one has
\begin{eqnarray}
&&\int^{\infty}_{-\infty}dy\,exp[-|x-y|/L]
exp[-|y-z|/L]\nonumber\\
&&=(L+|x-z|)exp[-|x-z|/L] .
\end{eqnarray}
This shows that a multiplication of two operators,
which decay exponentially, produces an operator which
decays with the same exponential factor up to a polynomial
prefactor. A generalization of this relation suggests
that a suitable $(2k+1)$-th root of an exponentially
decaying operator gives rise to an operator with
an identical localization length for any finite $k$.
This holds in the case of the free fermion operator.
Therefore we examined the locality of interacting 
$H_{(2k+1)}$, as a direct extension of
the analysis in the overlap Dirac 
operator\cite{HJL}\cite{N2}. For 
$H_{(2k+1)}$(\ref{gwo1}) we showed that
$\| D_W^{(2k+1)\dagger}D_W^{(2k+1)}\| >0$
for all finite $k$ when 
$\| 1-U_{\mu\nu}\| $ is very small.
This means $H_{(2k+1)}$ is exponentially local.
As for the localization domain for $k=1$, for example,
we see
\begin{eqnarray}
&&\| 1-U_{\mu\nu}\| < \f{1}{2\times 10^5}\longrightarrow
\| D_W^{(3)\dagger}D_W^{(3)}\| >0\nonumber
\end{eqnarray}
Since this analysis gives a conservative estimate, we
expect that the actual locality domain of gauge field  
strength could be much larger.
But the operators spreads over more lattice points
for larger $k$. 

\section{Summary}
We discussed the properties of the Dirac operators
satisfying the algebraically generalized 
Ginsparg-Wilson relation. Our analysis indicates 
an infinite number
of lattice Dirac operators which have good properties. 
For the future work it is important to see the locality
of interacting operator more precisely. 
%and 
%to perform the weak coupling perturbation theory. 


\begin{thebibliography}{99}
\bibitem{GW}
P.H. Ginsparg and K.G. Wilson, Phys. Rev. {\bf D25} (1982)2649.
\bibitem{N1}
H. Neuberger, Phys. Lett.{\bf 417B}(1998)141\\
;{\bf 427B}(1998)353.
\bibitem{H}
P.~Hasenfratz,
Nucl. Phys. {\bf B525} (1998) 401.
\bibitem{HLN} 
P. Hasenfratz, V. Laliena and F. Niedermayer, Phys. Lett. 
{\bf 427B}(1998)125.
\bibitem{L1}
M. L\"{u}scher, Phys. Lett. {\bf 428B}(1998)342.
\bibitem{HJL}
P. Hernandez, K. Jansen and M. L\"{u}scher, Nucl. Phys. {\bf B552}
(1999)363.
\bibitem{N2}
H. Neuberger, Phys.Rev.{\bf D61}(2000)085015.
\bibitem{F1}
K. Fujikawa, Nucl. Phys. {\bf B589}(2000)487.
\bibitem{FI1}
K. Fujikawa and M. Ishibashi, Nucl. Phys. {\bf B587}(2000)419.
\bibitem{FI2}
K. Fujikawa and M. Ishibashi, Nucl. Phys. {\bf B605}(2001) 365.
\bibitem{C}
T.W. Chiu, Nucl. Phys. {\bf B588} (2000) 400;
Nucl. Phys. Proc. Suppl.{\bf 94} (2001) 733.
\bibitem{K}
Y. Kikukawa, private communication.
\end{thebibliography}
\end{document}